\documentclass[aps,prl,twocolumn,showpacs,preprintnumbers,amsmath,superscriptaddress,amssymb,floats,nofootinbib]{revtex4}
\usepackage{graphicx}
\begin{document}

\def\Journal#1#2#3#4{{#1} {\bf #2}, #3 (#4)}
\def\NCA{\rm Nuovo Cimento}
\def\NPA{{\rm Nucl. Phys.} A}
\def\NIM{\rm Nucl. Instrum. Methods}
\def\NIMA{{\rm Nucl. Instrum. Methods} A}
\def\NPB{{\rm Nucl. Phys.} B}
\def\PLB{{\rm Phys. Lett.}  B}
\def\PRL{\rm Phys. Rev. Lett.}
\def\PRD{{\rm Phys. Rev.} D}
\def\PRC{{\rm Phys. Rev.} C}
\def\ZPC{{\rm Z. Phys.} C}
\def\JPG{{\rm J. Phys.} G}


\title{The search for $\phi$-N bound state from subthresold production 
of $\phi$ meson}

\author{S. Liska, H. Gao, W. Chen, X. Qian}
\affiliation{Department of Physics and the Triangle Universities Nuclear Laboratory, Duke University, Durham, NC 27708-0305}

\noaffiliation

\date{\today}

\begin{abstract}
 The subthreshold photoproduction of $\phi$ meson from heavy nuclear targets 
has been suggested as a candidate to search for the $\phi$-N bound state, 
a quantum chromodynamics (QCD) molecular state. In this paper, we present 
detailed Monte Carlo studies to demonstrate the feasibility of this technique.
Further, we show that proton induced subthreshold production of $\phi$ 
meson from heavy nuclear targets is also suitable for such a search.

\end{abstract}
\pacs{13.60.Le, 24.85.+p, 25.10.+s, 25.20.-x, 25.40.-h}
\maketitle

S. J. Brodsky, I. A. Schmidt and G. F. de T\'{e}ramond~\cite{Brodsky:1990jd} suggested 
that the QCD van der Waals interaction, mediated by multi-gluon exchanges, is dominant 
when the two interacting color singlet hadrons have no common quarks.
Luke, Manohar and Savage~\cite{Luke:1992tm} predicted that the QCD van der Waals interaction 
is enhanced at low velocity.
This finding supports the prediction that 
a nuclear-bound quarkonium can be produced in charm production 
reactions at threshold.  
Brodsky, Schmidt, and de T\'{e}ramond \cite{Brodsky:1990jd} investigated 
the nuclear-bound quarkonium state using
a non-relativistic Yukawa type attractive
potential $V_{(Q\bar{Q})A}= -\alpha e^{-\mu r}/r$
characterizing the QCD van der Waals interaction. They determined the 
$\alpha$ and $\mu$ constants using the phenomenological model of
high-energy Pomeron interactions~\cite{landshoff}. 
Using a variational wave function 
$\Psi(r) = (\gamma^3 /\pi)^{1/2}e^{-\gamma r}$, they predicted bound
states of $\eta_{c}$ with $^3$He and heavier nuclei. This prediction
was later confirmed by Wasson~\cite{wasson} using a more realistic 
$V_{(Q\bar{Q})A}$ potential taking into account the nucleon
distribution inside the nucleus.

Similarly, one expects the attractive QCD van
der Waals force dominates the $\phi$-N interaction 
since the $\phi$ meson is almost a pure $s\bar{s}$ state.
Using the variational method and following Ref.~\cite{Brodsky:1990jd} to
assume $V_{(s\bar{s}), N}=-\alpha e^{\mu r}/r$, 
Gao, Lee and Marinov~\cite{gao} find that a bound state of 
$\phi$-N is possible with $\alpha=1.25$   
and $\mu= 0.6$ GeV. The binding energy obtained is 1.8 MeV. 
Their results should be compared with $\alpha=0.6$ and
$\mu=0.6$ GeV determined in \cite{Brodsky:1990jd} for the 
$c\bar{c}$ quarkonium.
The interaction is expected to be enhanced
by $(m_c/m_s)^3$, i.e., $q\bar q$ separation cubed, from $c\bar{c}$
to $s\bar{s}$.
Since the radius of the $\phi$ meson 
is 0.4 fm \cite{povh} twice
the radius of the $J/\Psi$ meson, $\alpha = 1.25$ is a rather
conservative coupling
constant to use for the $\phi$-N interaction. 
Also, the interaction is expected to have longer range for 
the $\phi$-N system than that of the $c\bar{c}$-N interaction.
Thus, $\mu$ = 0.6 GeV used in the variational approach described above
is also conservative for the $\phi$-N interaction. 
  
Recently, the $\phi$-N bound state has been studied by Huang, Zhang 
and Yu~\cite{huang} using a chiral SU(3) quark model and the extended chiral
SU(3) quark model solving the Resonating Group Method (RGM) equation.
The model parameters used by the authors in this work 
provided good description of baryon bound
states, deuteron binding energy and NN scattering phases shifts in their 
previous work~\cite{huang1,huang2}. 
A $\phi$-N quasi-bound state with several MeV of binding energy
was predicted by the aforementioned extended chiral quark model 
plus channel coupling 
effect.

Such a $\phi$-N bound state could be 
formed~\cite{gao}  
inside a nucleus at the quasifree subthreshold photoproduction 
kinematics where the attractive force is expected to be 
enhanced. 
Recently, ``subthreshold'' $\phi$ meson photoproduction has been 
observed~\cite{g10phi} for the 
first time from a deuterium target at Jefferson Lab.
The experimental search for such a bound state would be a triple
coincidence detection of kinematically correlated $K^{+}$, $K^{-}$,
and proton in the final state from subthreshold production of
$\phi$ meson from heavy nuclear targets.
To identify clearly the signal of a $\phi$-N
bound state, one needs to understand the background contribution to the 
$p K^{+} K^{-}$ final state carefully.  
The dominant background contribution to the signal of interest are the 
quasifree subthreshold production of $\phi$ meson from a bound proton 
and the subsequent decay of $\phi$ into $K^+$ $K^-$ without the formation 
of a $\phi$-N bound state, and the direct production of $K^+$ $K^-$ from a 
bound proton inside the nucleus. 
Recently, we carried out a detailed Monte Carlo simulation of these processes. 
The Monte Carlo study shows that cuts on the momentum 
correlation between proton
and $K^{\pm}$, and on the invariant mass of the $K^+$, $K^-$ 
and proton system can clearly separate the signal of the decay of a 
$\phi$-N bound state from the backgrounds. 
Therefore, one can identify a bound $\phi$-N state experimentally 
using the aforementioned triple coincidence experimental technique.    
In this paper, we present our results from the Monte Carlo studies.

The kinematics for all three channels considered in the Monte Carlo study for 
the detection of a $\phi$-N bound state, using $^{12}$C, $^{56}$Fe, $^{63}$Cu 
and $^{197}$Au target nuclei and the CLAS detector~\cite{CNIM} 
at Jefferson Lab, 
follow a set of common parameters and assumptions. First, the energy of the 
photons in the region of interest is distributed uniformly from 1.65 to 1.75 
GeV. This energy range is roughly 80-180 MeV higher than that of 
a simulation in the case of an ideal detector
for the triple coincidence detection of proton, $K^+, K^-$. 
The target nuclei are assumed to be initially at rest in the lab
frame. For each event, the Fermi momentum and the missing energy for the 
bound proton are weighted by the nuclear spectral function. 
The nuclear spectral functions for the $^{56}$Fe, $^{63}$Cu, and $^{197}$Au 
nuclei used in our simulations are based on the Green's 
function Monte Carlo calculation~\cite{wiringa}, and from de-radiating 
data for the $^{12}$C target~\cite{dutta}. 
A Breit-Wigner distribution with a mean value of 1.019413 GeV/c$^2$ and a 
width 0.00443 GeV/c$^2$ is used to model the mass of the $\phi$ meson 
that is produced. Furthermore, the $\phi$ mass for each event is bound between 
0.819413 to 1.219413 GeV/c$^2$ to avoid physically unreasonable masses. 

Furthermore, the CLAS detector is modeled by applying cut-offs to the 
zenith lab angles that can be detected for the final state of each channel. 
From an analysis of the CLAS g10 data for the final state of
$p K^+ K^-$ from 
$\phi$ meson photoproduction~\cite{kramer}, the minimum zenith lab 
angles for $K^+$ and $K^-$ are 5 and 20 degrees, respectively.
Additionally, the resolution of the detector is incorporated into the 
simulation by weighing events using a Gaussian type distribution 
following the procedure
developed for the Jefferson Lab Hall C proposal on electroproduction of $\phi$ 
using Hall C SOS and HKS magnets employing the missing mass 
technique~\cite{gao2}. Realistic numbers for the CLAS detector resolutions are
used in our simulation.

The first channel considered is the $\phi$ meson production 
from a bound proton with the subsequent decay of $\phi$ into $K^+K^-$ without the formation of a $\phi$-N bound state. The events simulated are weighted 
by the mass of the $\phi$ meson, and the Fermi momentum and missing energy of the bound proton. Before computing the kinematics of $\gamma + ``p" \rightarrow p + \phi$, the energies of $\gamma$ and $``p"$ are checked to ensure they are 
sufficient to produce a $\phi$ meson; events with insufficient energy are 
discarded. Given that no bound state is formed, the kinematics of the $\phi$ 
meson decay into $K^+K^-$ are calculated. At this point in the simulation, events for which the zenith lab angle of $K^+$ or $K^-$ below the CLAS detector cut-offs are removed. Before simulating the detector's resolution, the calculations 
performed are tested in two ways: reconstructing the mass of the $\phi$ meson 
from the energy and linear momentum of the $p K^+ K^-$ final state; and 
reconstructing Fermi momentum of the bound proton from the 4-vector of the 
initial and final state particles. Finally, the detector resolution is 
simulated for the $p K^+ K^-$ final state. 

\begin{figure*}
\centering
\resizebox{4.5in}{!}{\includegraphics{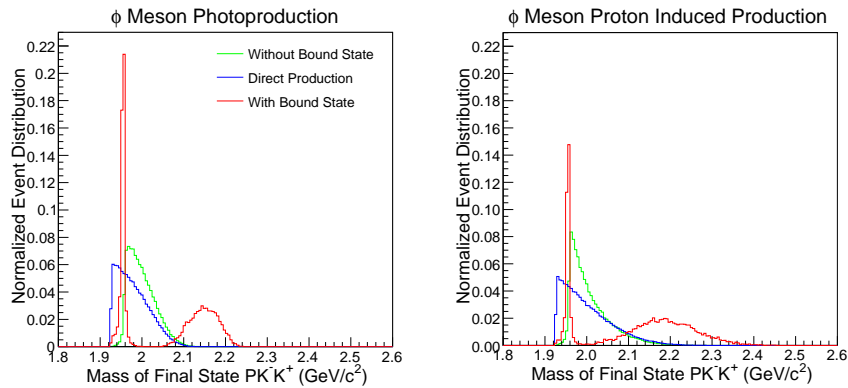}}
\caption{Monte Carlo simulation of the invariant mass distribution of the 
proton, $K^{-}$ and $K^{+}$ system 
for the following three processes from a copper target: $\phi$-N bound state 
(red); quasi-free subthreshold $\phi$ production without the formation of a
bound state (green); direct production of the $p K^+ K^-$ process (blue).  
The left and the right panel show the photo- and proton induced 
$\phi$ subthreshold production, respectively.}
\label{fig:invariantmass}
\end{figure*}

In the case of the second background channel, the direct production 
of $K^+K^-$ from a bound proton, the simulation's structure is essentially 
the same as the structure of the first channel considered. An important 
distinction for the second background channel is the direct production of the 
three-particle final state from the $\gamma + ``p" \rightarrow p + K^+ + K^-$ 
process, which is simulated by the following sequence of steps. 
For computational purposes, it is assumed that the direct production of 
$K^+K^-$ can be simulated by $\gamma + ``p" \rightarrow (AB) + C \rightarrow A+B+C$, where $A,B$ and $C$ are combinations of proton, $K^+$ and $K^-$. 
The intermediate step of the simulation, $(AB) + C$, has no physical 
significance, and merely serves as a tool for kinematic calculations. 
Given that $K^\pm$ are assumed to be kinematically indistinguishable in the 
simulation, there are only three kinematically distinct combinations, 
each of which is assumed to have an equal probability of occurring. Thus, 
each event is weighted by the uniform distribution of the different 
sequence of the same three-particle final state.
Similarly to the first channel, the Fermi momentum of the 
bound proton is reconstructed to ensure all calculations are carried out 
correctly. 

The last channel being studied is exactly the same as the first, with the 
exception of the formation of a $\phi$-N bound state in the nuclear medium. 
For the purposes of the simulation and the detection setup, the nucleon to 
which the $\phi$ meson binds is modeled as a proton. It is assumed that QCD van der Waals forces lead to the formation of a $\phi$-N between the 
photo-produced $\phi$ meson and a bound proton if the $\phi$ particle has a 
linear momentum less than 400 MeV/c. The momentum of the bound proton is 
assumed to be the same as that of the $\phi$ when considering the 
formation of the $\phi$-N state.
The $\phi$-N exotic state is assumed to have a binding 
energy 2.5 MeV~\cite{gao}. When considering the decay 
of $\phi$-N into the $p K^+ K^-$ final state, the kinematic methodology used 
for the direct production of $K^+K^-$ from bound proton is reused. Once again, 
the simulation is checked by reconstructing the mass of the $\phi$ meson 
and the Fermi momentum of the bound proton. 
Fig.~1 shows the invariant mass distribution of the $p, K^+, K^-$ system for 
all these three channels, and the histogram for each of the channels has been
normalized to unity. In the case of the $\phi$-N bound state, the prominent 
peak corresponds to the proton from the decay of the $\phi$-N bound 
state is detected with the second bump corresponding to the recoil proton(s)
 from the subthreshold $\phi$ production process being detected.

\begin{figure*}
\centering
\resizebox{4.5in}{!}{\includegraphics{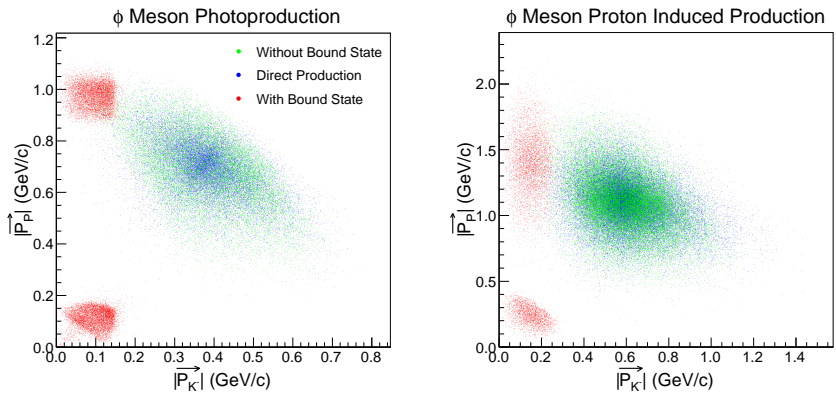}}
\caption{Monte Carlo simulation of the proton momentum versus $K^{-}$ momentum
for the three processes from a copper target described in the text. 
To improve the visibility of the figure, we
plot 25,000 successful events for each channel only, while our overall results
are based on analyzing 45,000 successful events for each channel.}
\label{fig:monte}
\end{figure*}

\begin{figure*}
\centering
\resizebox{4.0in}{!}{\includegraphics{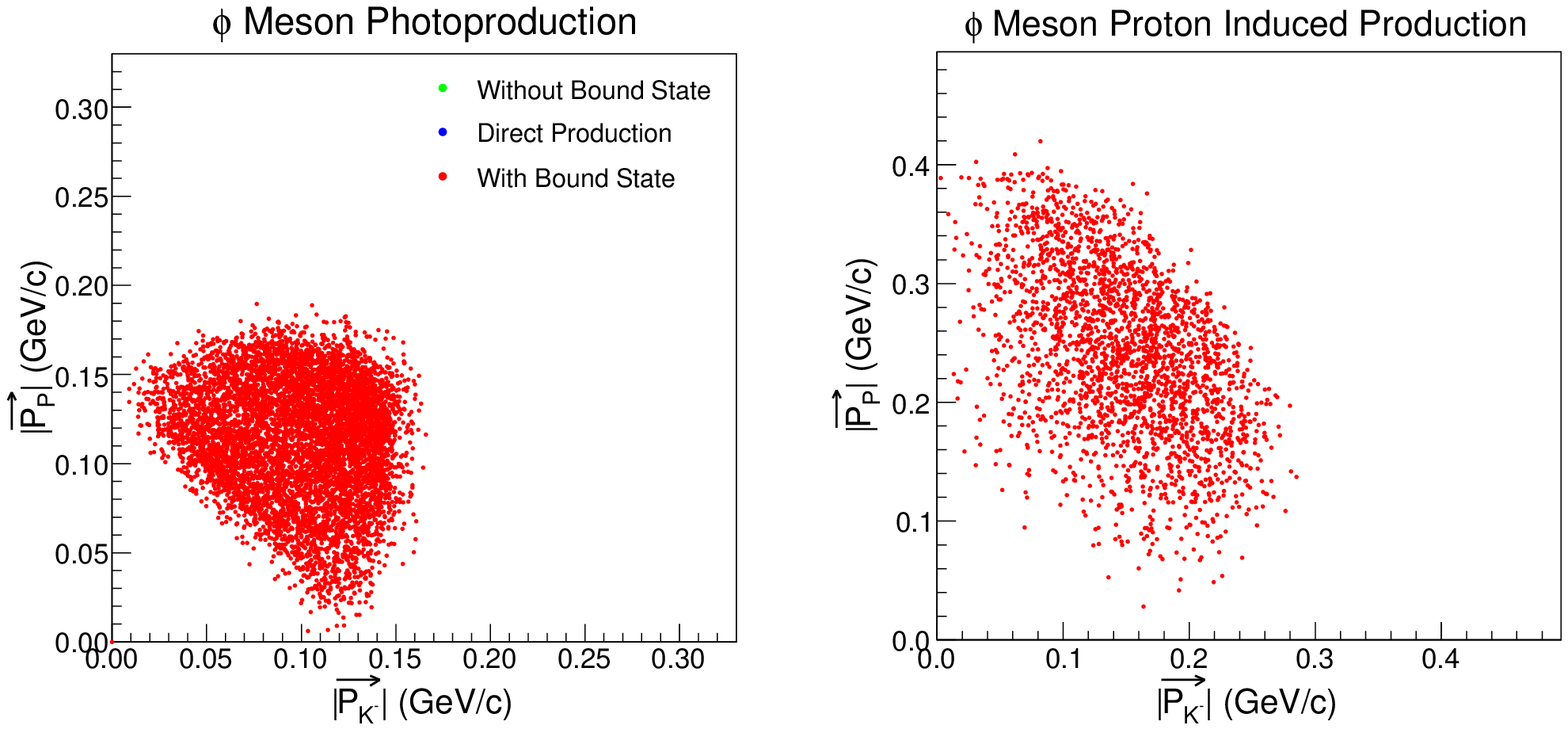}}
\caption{Same as Fig.~2 except for the momentum and invariant mass 
cuts applied (see text).}
\label{fig:monte_crop}
\end{figure*}

The search of the presence of a $\phi$-N bound state is based on the 
triple coincidence detection of the kinematically correlated $K^+$, $K^-$, 
and the proton in the final state. Such a signature is clearly identified 
when comparing the absolute value of the linear momentum of the $K^{\pm}$ 
and the scattered protons. We are able to segregate the events corresponding 
to the channel with the $\phi$-N bound state by applying one graphical cut, 
and another cut on the invariant mass of the $p K^+ K^-$ final state. The 
graphical cut involves removing all events for which the square sum of the 
linear momentum of $K^-$ or $K^+$ and the proton is greater than 
(0.300 GeV/c)$^2$. The second cut consists of only considering events where 
the invariant mass of the $p K^+ K^-$ final state falls in the region of 
$1.955 \pm 0.010$ GeV/c$^2$. The results from the simulation using $^{12}$C, 
$^{56}$Fe, $^{63}$Cu and $^{197}$Au target nuclei have demonstrated that the 
channel for $\phi$-N is completely segregated from the two other background 
channels. Only about 11\% of events for the channel containing the $\phi$-N 
bound state are eliminated due to these two cuts with no additional 
cuts from the detectors. 
An example 
is shown for the case of the Copper nucleus before (Fig.~2) and 
after (Fig.~3) the 
momentum and invariant mass cuts discussed in the text.

To increase the feasibility of experimental detection of a $\phi$-N bound
state a Monte Carlos analysis similar to that of subthreshold $\phi$ meson
photoproduction is carried out for subthreshold proton 
induced $\phi$ production. The
overall conclusions from the proton induced production are the same as those
for photoproduction, namely that the events containing $\phi$-N bound states
can be completely separated from background channels. The three channels
considered for the proton induced $\phi$ production are equivalent to those
considered in photoproduction, with the distinction of the different incident 
beam.
Once again the simulation is conducted in the subthreshold region, to
minimize the momentum mismatch between the produced $\phi$ meson and the bound
nucleons, and to obtain a clear kinematic signature for the presences of
$\phi$-N bound state when comparing the final state momentum value of the
final products. The incident proton's total energy for the simulated events
is weighed by a uniform distribution in the subthreshold range of 3.30 -
3.50 GeV, where the threshold energy is 3.53 GeV.

Even though the analysis for proton induced production is similar to that of
photoproduction; there are some important kinematic distinctions that arise
from having to consider an additional proton during the various stages of
the simulation. When simulating the channels where a $\phi$ meson is produced
from a bound proton the methodology used for the direct production of
$K^+ K^-$ in
photoproduction is reused noting that the different kinematic combinations
now involve two protons and one $\phi$ meson. For the channel corresponding to
the direct production of $K^{+}K^{-}$ we need to consider the four-particle 
final state:
$p + "p" = p + p + K^{+} + K^{-}$. This process is simulated by $p + "p"
\rightarrow (AB) + (CD) \rightarrow A + B + C + D$, where $A,B,C$, and $D$ are
combinations of $p,p, K^{+}$, and $K^{-}$. Following the assumption that
$K^{\pm}$ are kinematically indistinguishable we observed only two
dynamically different sequences, which are attributed an equal probability
of occurring. Each event is weighted by uniform distribution of the two
different sequences. Another important difference of the proton induced $\phi$
production when compared to photoproduction is that for the channel assuming
the presence of a $\phi$-N bound state the condition for such a bound state
occurring is that the linear momentum of the $\phi$ meson is less 500 MeV/c
(as oppose to the 400 MeV/c used for photoproduction). 
This increase of the 
$\phi$ meson upper momentum limit is necessary due to the overall higher 
momentum distribution of the $\phi$ meson produced. Although high momentum
protons inside nuclei are suppressed, one still expects more $\phi$-N bound 
state events 
from proton induced reaction due to the nature of the strong interaction. 
 Such an experiment can be carried out in the future
at the CSR facility in Lanzhou, China where one can design a new detection 
system particularly suitable for this search.

The presence of a $\phi$-N bound state is observed when comparing the final
state momentum of the protons and kaons using two cuts to isolate 
the channel with the $\phi$-N
bound state in the proton induced production. The first cut removes all the
events for which the squared sum of the final state momentum of
$K^-$ or $K^+$ and proton is greater than (0.450 GeV/c)$^{2}$. The second
cut is the same as that for the photoproduction case which is
on the invariant mass of the final state $pK^{-}K^{+}$.
These two cuts yield a clear and uncontaminated kinematic
signature for the presence of a $\phi$-N bound state. The percent on events
from the channel containing the $\phi$-N bound state eliminated due to these
two kinematic cuts is about 11\%  with no additional 
cuts from the detectors.  

In the two cases presented in this work, the kinematic 
cuts described are able to separate the signal cleanly from the
main physics backgrounds, therefore it is not necessary to carry out the 
simulations with cross-section weighting.
To carry out such a search at Jefferson Lab, the feasibility of using 
the CLAS BONUS~\cite{bonus} type recoil detector 
for detecting low-momentum charged particles
will be studied in the near future.

In summary, we carried out
detailed Monte Carlo studies of subthreshold $\phi$ meson production 
from photo- and proton induced reactions to demonstrate the feasibility of 
experimental search for $\phi$-N bound state from heavy nuclear targets at
Jefferson Lab in 
U.S. and the CSR facility in Lanzhou, China. 
This work is 
supported in part by the U.S. Department of Energy under contract
number DE-FG02-03ER41231. H.G. also acknowledges the support of the 
National Science Foundation of China through an overseas young scholar 
collaborative award.

\end{document}